\documentclass[aps,10pt,secnumarabic,twocolumn,amssymb,notitlepage,superscriptaddress]{revtex4-2}
\usepackage[utf8]{inputenc}
\usepackage[T1]{fontenc}
\usepackage{epsf,graphicx,amssymb,amsmath,color,float,subfig}
\usepackage[dvipsnames]{xcolor}
\usepackage[squaren, Gray, cdot]{SIunits}
\usepackage[english]{babel}
\usepackage{hyperref}
\hypersetup{colorlinks=true,linkcolor=blue,citecolor=blue}
\usepackage{orcidlink}
\usepackage[defaultcolor=red]{changes}

\newcommand{\phyp}{\textit{Physarum polycephalum}}

\begin{document}

\title{Slow modulation of the contraction patterns in \emph{Physarum polycephalum}}

\author{\orcidlink{0000-0001-6761-1162}{Raphael Saiseau}}\email{raphael.saiseau@uni-konstanz.de}
\affiliation{Laboratoire MSC, Universit\'{e} Paris Cité, CNRS, UMR 7057, Mati\`{e}re et Syst\`{e}mes Complexes (MSC), F-75006 Paris, France.}
\affiliation{Department of Physics, University of Konstanz, 78457 Konstanz, Germany}

\author{Valentin Busson}
\affiliation{Laboratoire MSC, Universit\'{e} Paris Cité, CNRS, UMR 7057, Mati\`{e}re et Syst\`{e}mes Complexes (MSC), F-75006 Paris, France.}

\author{\orcidlink{0000-0002-6619-1466}{Marc Durand}}\email{marc.durand@univ-paris-diderot.fr}
\affiliation{Laboratoire MSC, Universit\'{e} Paris Cité, CNRS, UMR 7057, Mati\`{e}re et Syst\`{e}mes Complexes (MSC), F-75006 Paris, France.}

\begin{abstract}
The slime mould \phyp~has emerged as a model for self-organisation and coordination of contractile activity at large spatial scales. This self-organisation largely results from cytoplasmic flows generated by propagating contractile waves of the actomyosin cortex. In addition to these relatively fast travelling waves, complex slow modulations of the contractile activity have been observed on timescales much longer than the primary oscillation period; these slow dynamics are however scarcely characterised.
Here we characterise these slow modulations by confining organisms inside annular geometries. 
We quantify contractile activity simultaneously across the entire organism on long time scales, exhibiting correlations between contractile wave direction, amplitude modulation, and the moving mean vein diameter. 
We observe travelling and alternating wave patterns: travelling wave periods scale clearly with system size, while alternating wave periods remain broadly distributed and probe larger values as the system size increases. Strikingly, the measured periods align with integer multiples of an intrinsic modulation time scale obtained independently from statistical analysis.
These observations support the hypothesis that transport of a slowly advected chemical agent, which locally modifies membrane/cortex mechanical properties, underpins the observed slow modulation dynamics, accounting for the coordination across the organism on long time scales.
\end{abstract}

\maketitle

\section*{Introduction}

Flows sustained over long distances are essential to many biological processes, from migration or development to the dissemination of resources and signals. 
A canonical mechanism to generate cytoplasmic flows is coordinated mechanical contraction-extension of the actomyosin cortex, where locally the periodic contraction pumps the internal fluid~\cite{grebecki1978correlation,grkebecki1978dynamics}. 
Chemical species (for example calcium or cAMP) regulate contractile activity~\cite{kscheschinski2023calcium,kohama1987inhibitory}, and their advection by flow provides a plausible route for non-local communication. At the same time mechanical stresses produced by the flow and by cortex deformations feed back on contraction activity through mechanosensitive pathways, so that mechano-chemical coupling is expected to enable this long-range coordinated activity~\cite{radszuweit2013intracellular,julien2018oscillatory,staddon2022pulsatile,werner2024mechanical,picardo2025active}. 
External forces~\cite{umedachi2017response}, environmental factors~\cite{etienne2015cells}, local damage~\cite{bauerle2017spatial} or nutrient intake~\cite{kramar2021encoding} are known to strongly influence contractile patterns and the architecture of the supporting filamentous network. This mechano-chemical mechanism then serves as a strong candidate framework behind organism-scale coordination, organizing internal activity and enabling global responses to local stimuli.

\phyp\ is a large, multinucleated unicellular organism that forms a centimetre-scale transport network. Conspicuous, regularly organised, contractile oscillations of the actomyosin cortex drive macroscopic cytoplasmic flows throughout its network. 
The dominant dynamics are travelling waves with a typical time period of order $10^{2}$~s and a wavelength comparable to the organism size \cite{busson2022emergence, alim2013random}. 
These fast waves have been the subject of numerous studies, characterising their travelling nature~\cite{fleig2022emergence} and spatial patterns~\cite{busson2022emergence}, relating them to organism size and geometry~\cite{kuroda2015allometry}, and investigating their relation to migration~\cite{oettmeier2019lumped}.

\begin{figure}[t]
\centering
\includegraphics[width=\columnwidth]{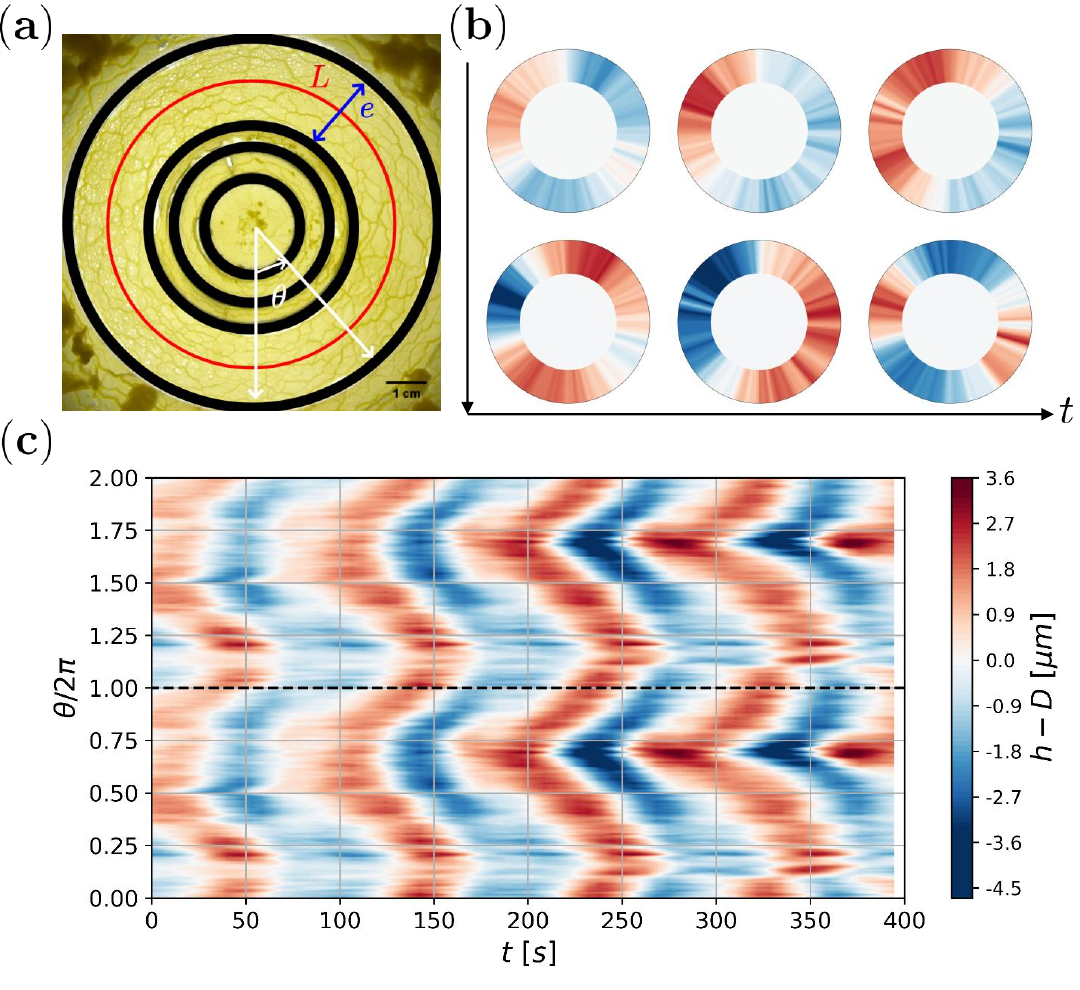}
\caption{(a) Bright-field raw image of three concentric ring-confined plasmodia. The main-axis perimeter $L$ and ring width $e$ are indicated. (b) Height/thickness variation along the ring at successive times; counter-propagative waves are visible. (c) Space-time plot (kymograph) of the contractile oscillations $h(\theta,t)-D(\theta,t)$ (see signal decomposition Eq.~\eqref{Eq:h_def}), revealing the fast travelling phase waves.}
\label{fig:Kymo_planewave}
\end{figure}

\begin{figure*}[t!]
\centering
\includegraphics[width=\textwidth]{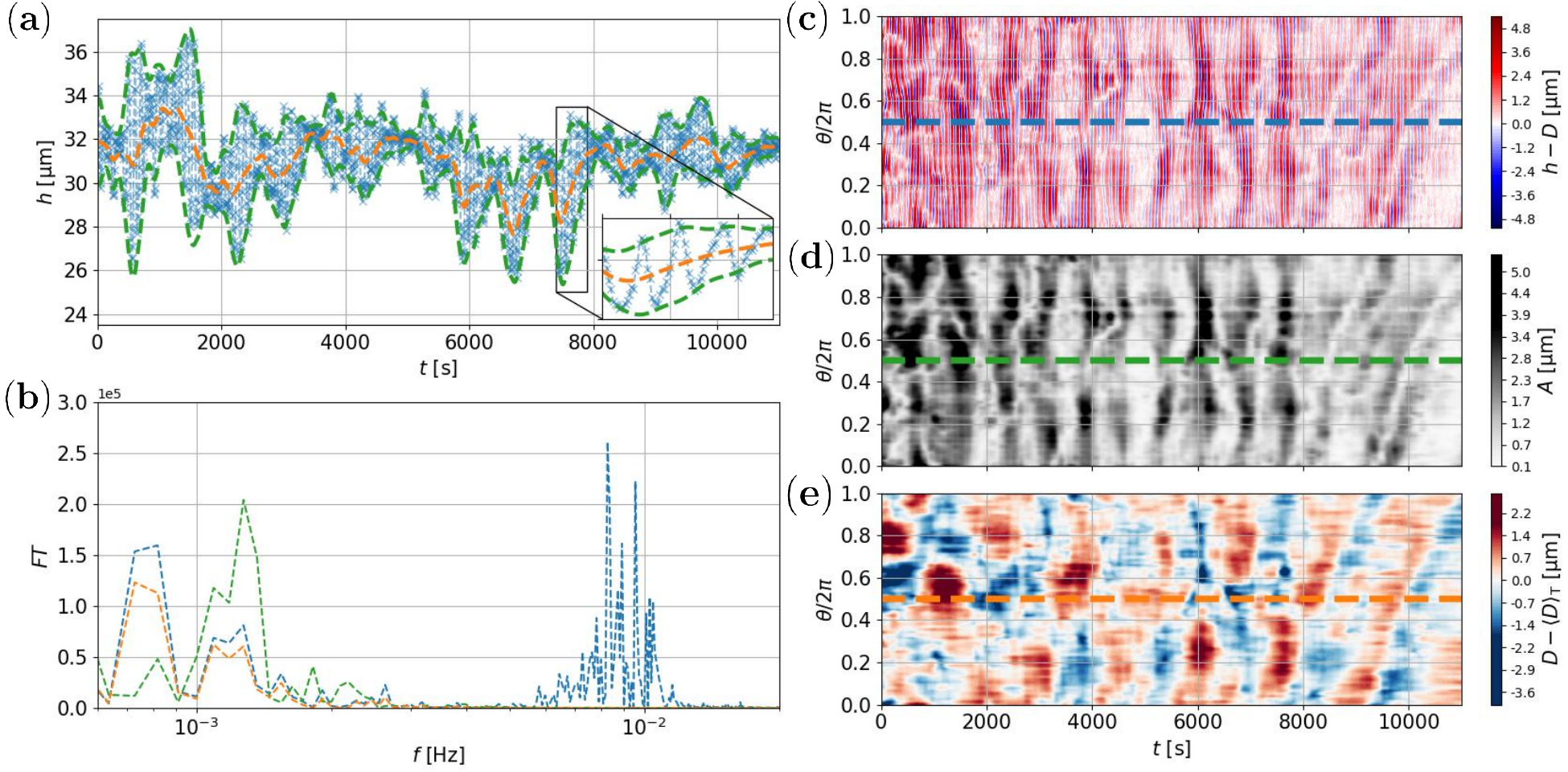}
\caption{(a) Example time series of $h(\theta=\pi,t)$ (blue) with the extracted slow drift $D$ (orange) and amplitude envelope $A$ (green) defined by Eq.~\eqref{Eq:h_def}. (b) Fourier spectra of $h$, $A$ and $D$ for the same angular position. Kymographs of (c) the short-time oscillations $h(\theta,t)-D(\theta,t)$ displaying the phase waves, of (d) the amplitude modulation $A(\theta,t)$, and (e) the drift variations $\Delta D= D(\theta,t)-\langle D\rangle_T$.}
\label{fig:Kymo_long_comparison}
\end{figure*}

In contrast, slow modulations of the contractile waves at the organism scale, including amplitude modulation, long-term changes in mean vein diameter, and systematic variations in wave propagation direction, have received little focused attention~\cite{alim2017mechanism}. These slow dynamics, occurring on time scales of $\sim15$--$40$~minutes ($\sim10^{3}$s), have been reported in a variety of experimental contexts, from spontaneous long-time variations \cite{oettmeier2019lumped, busson2022emergence, kuroda2015allometry, fleig2022emergence, mayne2017coupled, marbach2023vein} to responses elicited by stimuli or damage \cite{kramar2021encoding, nakagaki2000interaction, yoshimoto1981simultaneous, bauerle2017spatial}. The ubiquity of these slow modulations suggests a general, possibly transport-mediated mechano-chemical mechanism coordinating activity across the network~\cite{julien2018oscillatory}, making \phyp\ an excellent candidate for studying this coordination through slow modulations.

Here we systematically investigate the contractile activity by confining \phyp\ in annular geometries (see Figure~\ref{fig:Kymo_planewave}(a)), as done previously for contraction oscillations~\cite{busson2022emergence} and for network reorganisation dynamics~\cite{saiseau2024network}. This geometry simplifies boundary conditions (no free polarity axis) and reduces the problem to a quasi-1D geometry defined by angular position $\theta$. We decompose the thickness signal into a drift, an amplitude modulation and a phase, and analyse spatial patterns, temporal scales and scaling with system size. 
Using this geometry and decomposition, we previously revealed that fast-phase spatial patterns strikingly exhibit a dominant counter-propagative wave~\cite{busson2022emergence}.  Figure~\ref{fig:Kymo_planewave} gives a representative example of such an oscillatory wave with a period of $100$~s, exhibiting clear source and sink locations near the top-left and bottom-right of the ring; panel (b) shows the projection onto the confining ring and panel (c) the corresponding kymograph. Here we focus our attention to the slower modulation dynamics.
Our experiments reveal that amplitude modulation, mean vein diameter changes, and directionality of phase waves are strongly correlated in space and time.
These slowly varying fields exhibit rotating as well as standing/alternating modes (despite the later more scarcely), and a systematic spectral analysis identifies a robust intrinsic modulation timescale. The global mode periods scale with system size in a manner consistent with constant-speed transport ($T\propto L$) and agree with independently reported modulation and transport velocities \cite{alim2017mechanism}. 

Altogether, these observations support the hypothesis of a slowly varying regulator, plausibly a chemical messenger, whose transport by cytoplasmic flows modulates local excitability and mechanical properties. Variation of this advected field across the network would then bias local oscillator dynamics and mediate long-range coordination, ultimately producing the observed spatio-temporal mode formation.

\section*{Materials and Methods}

Homogeneous plasmodia were prepared following standard culturing protocols (see Supplementary Materials for culture and insemination methods). We confined plasmodia in concentric annular chambers (rings) formed by cylindrical walls pressed into the growth substrate; ring widths and radii were chosen to produce aspect ratios ranging from 11 to 41. Each ring is characterised by its main-axis radius $R$, perimeter $L=2\pi R$, and width $e$ (see Fig.~\ref{fig:Kymo_planewave}a).

Rings were imaged in transmission using a colour digital camera. We extracted the blue channel to maximise contrast between plasmodium and background and used an empirically calibrated conversion from transmitted intensity to a thickness proxy $h(t,\theta)$ (see Supplementary Materials for image acquisition, preprocessing and calibration details). Images were mapped to polar coordinates and resampled to obtain $h$ as a function of time $t$ and angular position $\theta\in[0,2\pi]$.

In total we analysed 47 ring experiments (perimeters $L$ ranging from 6.0 to 13.5 cm), corresponding to 16 different specimen, with a typical recording length per ring ranging from 4 to 12 hours.
To characterise the multiple time scales we decompose the local thickness as:
\begin{equation}
h(t,\theta) = D(t,\theta) + A(t,\theta)\cos\left[\phi(t,\theta)\right],
\label{Eq:h_def}
\end{equation}
where $\phi(t,\theta)$ is the fast phase (order $10^{2}$s), $A(t,\theta)$ the slowly varying oscillation amplitude, and $D(t,\theta)$ a slow drift capturing persistent vein diameter or density changes (due to network coarsening). 

Because network reorganisation occurs on $\sim10^{4}$~s timescale, we compute the drift moving average $\langle D\rangle_T$ averaged over the moving time window $T=3000\,$s to isolate vein-diameter dynamics from still slower network structural changes. Phase and amplitude are extracted by bandpass filtering around the primary oscillation and computing the analytic signal via the Hilbert transform (see Supplementary Methods for decomposition details). 

\section*{Results}

\subsection*{Slow modulation signal and separation of time scales}

Figure~\ref{fig:Kymo_long_comparison} shows a representative example of the raw thickness signal and its decomposition into drift, amplitude and phase (Eq.~\eqref{Eq:h_def}). The Fourier spectra (Fig.~\ref{fig:Kymo_long_comparison}(b) reveal a clear separation of time scales: the phase $\phi$ contains the fast component at $\sim10^{2}$\,s, while the amplitude $A$ and drift $D$ evolve on a slower timescale $\sim10^{3}$\,s. A striking empirical observation in these data is that the slowly varying amplitude envelope $A(t,\theta)$ closely tracks the absolute value of the slow drift variations $|\Delta D|= |D(t,\theta)-\langle D\rangle_T|$ (see Fig.~\ref{fig:Kymo_long_comparison}(a)): peaks in $A$ co-locate with the magnitude extrema of $D$ in Fig.~\ref{fig:Kymo_long_comparison}(b). Because taking the absolute value of a slowly sign-changing signal doubles its spectral frequency (or, equivalently, halves its period), the relation $A\approx|\Delta D|$ has a simple and testable spectral consequence: if the amplitude spectrum shows a dominant period $T$, then the drift is expected to display a dominant period near $2T$. We return to this point below in the statistical analysis when comparing intrinsic modulation periods extracted from $A$ and from $D$.


\subsection*{Spatial patterns characterisation}

Spatially, $A$ and $\Delta D$ display both travelling (rotating) and alternating patterns. Across our dataset, distinct slow spatio-temporal patterns in $A$ and $\Delta D$ account for $38\%$ of the total analysed time segments (see \emph{Pattern classification} in Supplementary Methods). These patterns group broadly into: (i) \emph{rotating} modes, where amplitude or drift patterns steadily travel around the ring (including simple and counter-rotating travelling waves and higher-order harmonics); and (ii) \emph{alternating} modes, exhibiting fixed nodes and antinodes in angular space with oscillatory amplitude but no net rotation. 
These spatio-temporal patterns are observed more frequently and prominently in rings with larger radii and widths. Conversely, they cannot be seen in rings that are dying locally or have been cut (see Supplementary Figures~S2 and~S3). 
In Fig.~\ref{fig:Kymo_long_comparison} we highlight a transition between a rotating and an alternating mode to emphasize the often sudden nature of such switches.


Rotating patterns (i) predominate, accounting for $88\%$ of the observed patterns. Among these rotating patterns, single-wave fundamental modes are the most common (roughly $60\%$ of rotating-mode time), with harmonics and counter-rotating patches forming the remainder. Rotating episodes are particularly stable with a mean duration of $1.00\times 10^{4}$s (std.\ dev.\ $0.51\times 10^{4}$s) and some episodes exceeding $2.0\times 10^{4}$s. They can also reverse direction abruptly (Supplementary Fig.~S1). This time scale matches the network growth and reorganisation timescale previously reported for these experiments~\cite{saiseau2024network}, supporting the interpretation that rotating episodes are stable modes in the absence of architectural change.

\begin{figure}
\includegraphics[width=\columnwidth]{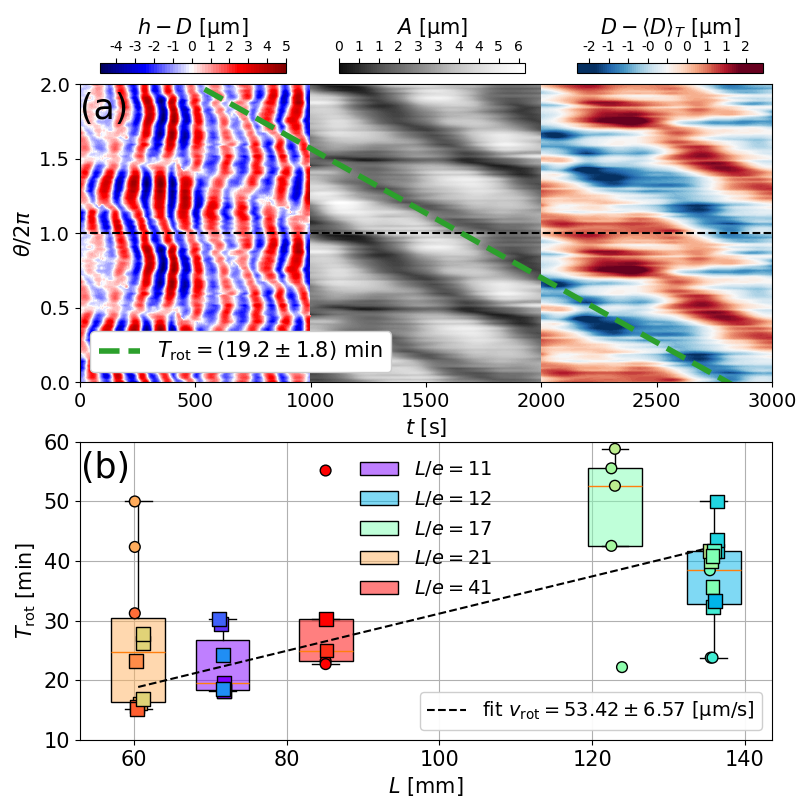}
\caption{(a) Example kymograph of the short-time height ($h-D$), amplitude $A$, and drift variations $\Delta D =D-\langle D\rangle_T$ showing a rotating pattern. The green dashed line is a linear fit of an isophase used to extract $T_{\mathrm{rot}}$. (b) Revolution period $T_{\mathrm{rot}}$ as a function of ring perimeter $L$ for multiple rings. Square markers: fundamental rotating modes. Round markers: harmonics or counter-rotating patterns. Colors indicate ring aspect ratio. Linear fit is shown with a black dashed line.}
\label{fig:Rot_pattern}
\end{figure}

We extract the revolution period $T_{\mathrm{rot}}$ by linear fits of isophase lines in $A(\theta,t)$ kymographs (Fig.~\ref{fig:Rot_pattern}(a)). Fig.~\ref{fig:Rot_pattern}(b) shows $T_{\mathrm{rot}}$ as a function of ring perimeter $L$ for multiple rings. The fitted relation is consistent with $T_{\mathrm{rot}}\propto L$, corresponding to an approximately constant transport speed of the amplitude/drift signal. A linear fit yields
\[
v_{\mathrm{rot}}=53.4\pm 6.6~\mathrm{\mu m/s}\,,
\]
based on 38 observed patterns. Note that harmonics or counter-rotating configurations (round markers in Fig.~\ref{fig:Rot_pattern}(b)) regularly deviate from the fundamental linear trend.

\begin{figure}
\centering
\subfloat[]{
\includegraphics[width=0.48\columnwidth]{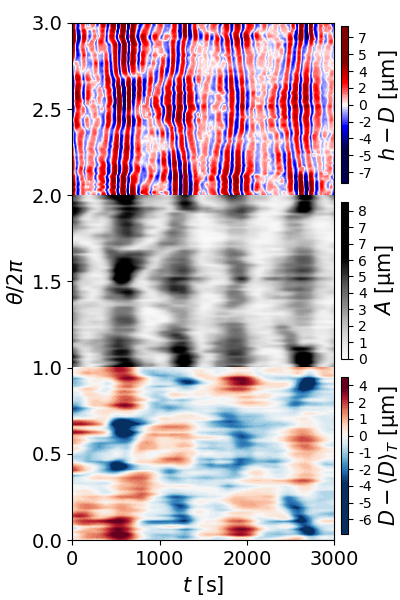}
}
\subfloat[]{
\includegraphics[width=0.49\columnwidth]{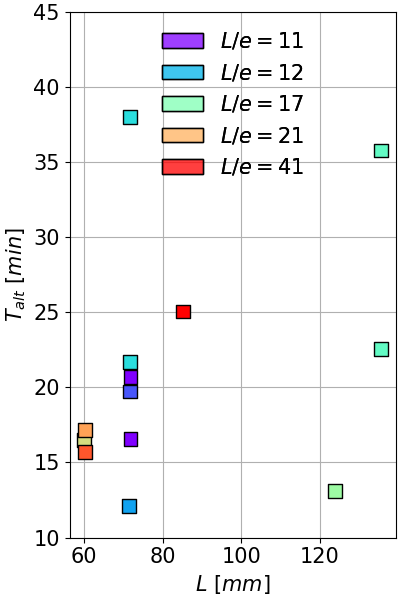}
}
\caption{(a) Alternating pattern period $T_{\rm alt}$ as a function of ring perimeter $L$. (b) Example kymographs of $h-D$, $A$, and $\Delta D =D-\langle D\rangle_T$ showing an alternating pattern.}
\label{fig:osc_direc}
\end{figure}

Alternating episodes (ii) are less common and typically observed as transitions, often between two rotating modes of opposite direction. They are shorter with a mean duration of $0.51\times 10^{4}$s (std.\ dev.\ $0.21\times 10^{4}$s). As the signal magnitude $|\Delta D|$ and hence $A$ often present two extrema per underlying drift cycle, we measure the period $T_{\rm alt}$ from the slowest signal $\Delta D$. 
The order of magnitude of the alternating time period $T_{\rm alt}$ is $10^{3}$s but shows substantial variability; its mean value typically increases with the system size (Fig.~\ref{fig:osc_direc}(a)). \\

Comparing the slow modulation patterns with the fast contractile oscillations, we find that the direction of the counter-propagative phase waves correlates with the gradient of $\Delta D$. The source and sink positions of the travelling phase waves (see Fig.~\ref{fig:Kymo_planewave}) define a polarity axis that rotates or alternates following the slow modulation patterns. The contraction/extension asymmetry is also frequently aligned with the drift pattern (compare Fig.~\ref{fig:Rot_pattern}(a) and Fig.~\ref{fig:osc_direc}(b)). Such patterns are most often present in networks that retain a well-developed central vein (Supplementary Fig.~S4).


\subsection*{Statistical analysis of modulation frequencies}

\begin{figure}
\centering
\subfloat[]{
\includegraphics[width=0.48\columnwidth]{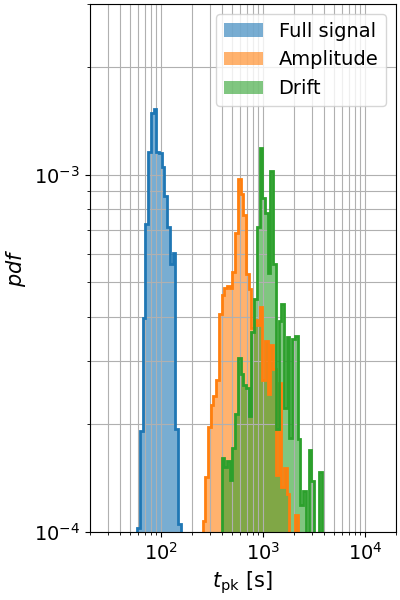}
}
\subfloat[]{
\includegraphics[width=0.48\columnwidth]{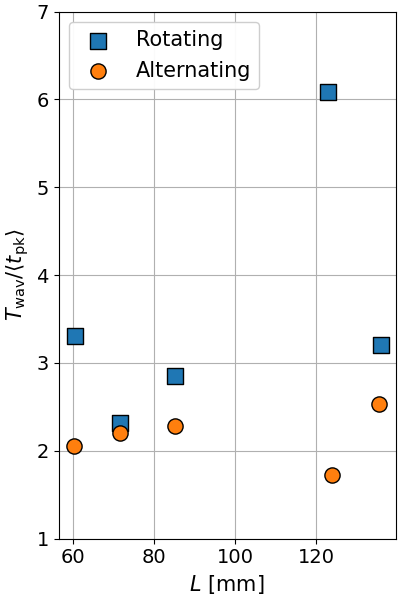}
}
\caption{(a) Weighted density histograms of the peak locations in the Fourier spectra of $h$, $A$ and $D$ for rings with $L=70$~mm. Weighting uses the peak power at each angular position. (b) Comparison of rotating and alternating pattern periods to the statistically most probable modulation period extracted from $A$ spectra.}
\label{fig:Fourier_spectrum}
\end{figure}

To identify an intrinsic modulation timescale, we compute Fourier spectra of $A(\theta,t)$ at each angular position for a given experiment, extract the frequency of the maximum peak and weight it by the corresponding peak power. 
Concatenating these values over all $\theta$ yields a weighted distribution of dominant frequencies for each experiment. Figure~\ref{fig:Fourier_spectrum}(a) shows the mean weighted distribution for experiments with ring perimeter $L=70$~mm, where a distinct peak is observed at $\langle t_{\rm pk}\rangle\approx 580$~s.
Because $A$ closely approximates $|\Delta D|$, the drift variation $\Delta D$ presents a spectral peak at approximately twice that period (i.e.\ near $1000\,$s), which then decreases rapidly for $t_{\rm pk} \gtrsim 3000~$s resulting from the signal processing. 

Figure~\ref{fig:Fourier_spectrum}(b) compares the statistically most probable modulation period (extracted from $A$) with the measured pattern periods. Consistent with the $A\approx|\Delta D|$ relation, alternating-pattern periods cluster near $2\langle t_{\rm pk}\rangle$ (the expected drift timescale), while rotating-mode revolution periods are observed at integer multiples of the intrinsic $\langle t_{\rm pk}\rangle$ as modes synchronise (in our dataset rotating-mode revolutions frequently appear near $3\times \langle t_{\rm pk}\rangle$).



\section*{Discussion and conclusion}

We report a systematic characterisation of slow modulation in contractile activity of \phyp\ confined to ring geometries. The ring geometry avoids polarity ambiguities and simplifies analysis by reducing the problem to a single angular coordinate, and by removing the network complex and evolving architecture from the analysis. 
Using a decomposition of thickness into phase, amplitude and drift, we find a clear separation of time scales between fast phase dynamics ($\sim10^{2}$s) and slow amplitude/drift modulation ($\sim10^{3}$s). 
Strikingly, the slowly varying amplitude field $A(\mathbf{x},t)$, the drift variations $\Delta D(\mathbf{x},t)$ (interpreted as mean vein diameter changes), and the source-sink locations of the fast phase waves are correlated in space and time.
Peaks in amplitude (markers of enhanced local contractile excitability) coincide with local extrema of the drift. Respectively, phase source and sink locations align with instantaneous drift maxima and minima. 
These empirical coincidences point to a slowly varying, system-scale regulator that simultaneously modulates excitability and the local mechanical state.

A natural candidate for this regulator is a chemical messenger (for example Ca\(^{2+}\)), advected by cytoplasmic flow and postulated to influence both contractility and cortical mechanics~\cite{alim2017mechanism, kscheschinski2023calcium}. In such a scenario the advected field sets a local amplitude \(A\) (via biochemical coupling) and alters local transport mobility or compliance (via mechanical softening), thereby biasing the local phase dynamics and enforcing a preferred direction of propagation~\cite{kramar2021encoding, zhang2017self}. \\

The slow modulation gives rise to robust spatio-temporal structures. The dominant behaviour are rotating modes, typically a single travelling wave that propagates with an approximately constant transport speed and can persist for long durations. Standing/alternating modes are also observed, despite less frequent that often act as transient regimes between rotating modes. Many global mode periods often align with integer multiples of a statistically identified intrinsic modulation time (of order $10^{3}\,$s), consistent with a timescales range given for local vein adaptation dynamics~\cite{marbach2023vein}.

Taken together, these observations favour an advection-dominated mechanism for the rotating modes: a slowly advected regulator biases local oscillators and promotes phase-locking across the ring. 
Advection implies a linear scaling between period $T$ and system size $L$ (\(T\!\propto\!L/v\)) for rotating modes, with a modulation transport speed matching the net material transport speed measured independently by PIV analysis~\cite{alim2017mechanism}.
Accordingly, the Péclet numbers estimated from typical material diffusivities (\(D_0\sim 10^{-10}\,\mathrm{m^2s^{-1}}\)) are large (\(\mathrm{Pe} = v_{\rm rot} L /D_0 \gg\)1), indicating that advection strongly dominates over molecular diffusion in the parameter range of our experiments (the same conclusion remains when estimating Taylor-enhanced diffusivity for veins of radius \(a\!\sim\!100\,\mu\)m).
In the coupled-oscillators picture, advection creates a spatial phase gradient so that after \(n\) local cycles the accumulated phase wind-up equals \(2\pi\), producing mode locking and a coherent rotating pattern \cite{Pikovsky_Rosenblum_Kurths_2001}.
Standing/alternating modes may then arise transiently when the mean advected flow reverses, or results from mode competition. Incidently, perturbations that stop advection hinder coordination between amplitude, diameter and phase-source locations (Supplementary Figures~S2 and~S3). 
Lastly, travelling and standing waves spontaneously emerge for large Peclet numbers for an active poroelastic gel with stress modulated by some chemical field~\cite{radszuweit2013intracellular}. \\

Finally, while these conclusions remain phenomenological and ask for targeted biochemical experiments to identify the responsible chemical agent, they hint towards a promising, low-cost method to infer the spatial map of rigidity and contractile activity across a network by tracking long-term vein-size evolution~\cite{wilkinson2023flow}. These results then indicate that the slow signal modulation is a promising observable to uncover the mechanism that coordinates flows and transport across the organism~\cite{nishikawa2017controlling}. 

Remarkably, whereas the fast phase waves break the ring's rotational symmetry by fixing source and sink locations, the slow modulation makes these locations drift and sample the entire circumference so that the time-averaged activity restore, over long times, the ring symmetry, further highlighting their central role in large-scale coordination.

\acknowledgments{
We acknowledge financial support from French National Research Agency Grants ANR-17-CE02-0019-01-SMARTCELL and CNRS MITI ‘Mission pour les initiatives transverses et interdisciplinaires’ (reference: BioRes).
}

\bibliography{biblio_Amplitude}

\begin{thebibliography}{28}%
\makeatletter
\providecommand \@ifxundefined [1]{%
 \@ifx{#1\undefined}
}%
\providecommand \@ifnum [1]{%
 \ifnum #1\expandafter \@firstoftwo
 \else \expandafter \@secondoftwo
 \fi
}%
\providecommand \@ifx [1]{%
 \ifx #1\expandafter \@firstoftwo
 \else \expandafter \@secondoftwo
 \fi
}%
\providecommand \natexlab [1]{#1}%
\providecommand \enquote  [1]{``#1''}%
\providecommand \bibnamefont  [1]{#1}%
\providecommand \bibfnamefont [1]{#1}%
\providecommand \citenamefont [1]{#1}%
\providecommand \href@noop [0]{\@secondoftwo}%
\providecommand \href [0]{\begingroup \@sanitize@url \@href}%
\providecommand \@href[1]{\@@startlink{#1}\@@href}%
\providecommand \@@href[1]{\endgroup#1\@@endlink}%
\providecommand \@sanitize@url [0]{\catcode `\\12\catcode `\$12\catcode `\&12\catcode `\#12\catcode `\^12\catcode `\_12\catcode `\%12\relax}%
\providecommand \@@startlink[1]{}%
\providecommand \@@endlink[0]{}%
\providecommand \url  [0]{\begingroup\@sanitize@url \@url }%
\providecommand \@url [1]{\endgroup\@href {#1}{\urlprefix }}%
\providecommand \urlprefix  [0]{URL }%
\providecommand \Eprint [0]{\href }%
\providecommand \doibase [0]{https://doi.org/}%
\providecommand \selectlanguage [0]{\@gobble}%
\providecommand \bibinfo  [0]{\@secondoftwo}%
\providecommand \bibfield  [0]{\@secondoftwo}%
\providecommand \translation [1]{[#1]}%
\providecommand \BibitemOpen [0]{}%
\providecommand \bibitemStop [0]{}%
\providecommand \bibitemNoStop [0]{.\EOS\space}%
\providecommand \EOS [0]{\spacefactor3000\relax}%
\providecommand \BibitemShut  [1]{\csname bibitem#1\endcsname}%
\let\auto@bib@innerbib\@empty
\bibitem [{\citenamefont {Grebecki}\ and\ \citenamefont {Moczo{\'n}}(1978)}]{grebecki1978correlation}%
  \BibitemOpen
  \bibfield  {author} {\bibinfo {author} {\bibfnamefont {A.}~\bibnamefont {Grebecki}}\ and\ \bibinfo {author} {\bibfnamefont {M.}~\bibnamefont {Moczo{\'n}}},\ }\bibfield  {title} {\bibinfo {title} {Correlation of contractile activity and of streaming direction between branching veins of physarum polycephalum plasmodium},\ }\href@noop {} {\bibfield  {journal} {\bibinfo  {journal} {Protoplasma}\ }\textbf {\bibinfo {volume} {97}},\ \bibinfo {pages} {153} (\bibinfo {year} {1978})}\BibitemShut {NoStop}%
\bibitem [{\citenamefont {Gr{\k{e}}becki}\ and\ \citenamefont {Cie{\'s}lawska}(1978)}]{grkebecki1978dynamics}%
  \BibitemOpen
  \bibfield  {author} {\bibinfo {author} {\bibfnamefont {A.}~\bibnamefont {Gr{\k{e}}becki}}\ and\ \bibinfo {author} {\bibfnamefont {M.}~\bibnamefont {Cie{\'s}lawska}},\ }\bibfield  {title} {\bibinfo {title} {Dynamics of the ectoplasmic walls during pulsation of plasmodial veins of physarum polycephalum},\ }\href@noop {} {\bibfield  {journal} {\bibinfo  {journal} {Protoplasma}\ }\textbf {\bibinfo {volume} {97}},\ \bibinfo {pages} {365} (\bibinfo {year} {1978})}\BibitemShut {NoStop}%
\bibitem [{\citenamefont {Kscheschinski}\ \emph {et~al.}(2023)\citenamefont {Kscheschinski}, \citenamefont {Kramar},\ and\ \citenamefont {Alim}}]{kscheschinski2023calcium}%
  \BibitemOpen
  \bibfield  {author} {\bibinfo {author} {\bibfnamefont {B.}~\bibnamefont {Kscheschinski}}, \bibinfo {author} {\bibfnamefont {M.}~\bibnamefont {Kramar}},\ and\ \bibinfo {author} {\bibfnamefont {K.}~\bibnamefont {Alim}},\ }\bibfield  {title} {\bibinfo {title} {Calcium regulates cortex contraction in physarum polycephalum},\ }\href@noop {} {\bibfield  {journal} {\bibinfo  {journal} {Physical Biology}\ }\textbf {\bibinfo {volume} {21}},\ \bibinfo {pages} {016001} (\bibinfo {year} {2023})}\BibitemShut {NoStop}%
\bibitem [{\citenamefont {Kohama}(1987)}]{kohama1987inhibitory}%
  \BibitemOpen
  \bibfield  {author} {\bibinfo {author} {\bibfnamefont {K.}~\bibnamefont {Kohama}},\ }\bibfield  {title} {\bibinfo {title} {Ca-inhibitory myosins: their structure and function},\ }\href@noop {} {\bibfield  {journal} {\bibinfo  {journal} {Advances in biophysics}\ }\textbf {\bibinfo {volume} {23}},\ \bibinfo {pages} {149} (\bibinfo {year} {1987})}\BibitemShut {NoStop}%
\bibitem [{\citenamefont {Radszuweit}\ \emph {et~al.}(2013)\citenamefont {Radszuweit}, \citenamefont {Alonso}, \citenamefont {Engel},\ and\ \citenamefont {B{\"a}r}}]{radszuweit2013intracellular}%
  \BibitemOpen
  \bibfield  {author} {\bibinfo {author} {\bibfnamefont {M.}~\bibnamefont {Radszuweit}}, \bibinfo {author} {\bibfnamefont {S.}~\bibnamefont {Alonso}}, \bibinfo {author} {\bibfnamefont {H.}~\bibnamefont {Engel}},\ and\ \bibinfo {author} {\bibfnamefont {M.}~\bibnamefont {B{\"a}r}},\ }\bibfield  {title} {\bibinfo {title} {Intracellular mechanochemical waves in an active poroelastic model},\ }\href@noop {} {\bibfield  {journal} {\bibinfo  {journal} {Physical review letters}\ }\textbf {\bibinfo {volume} {110}},\ \bibinfo {pages} {138102} (\bibinfo {year} {2013})}\BibitemShut {NoStop}%
\bibitem [{\citenamefont {Julien}\ and\ \citenamefont {Alim}(2018)}]{julien2018oscillatory}%
  \BibitemOpen
  \bibfield  {author} {\bibinfo {author} {\bibfnamefont {J.-D.}\ \bibnamefont {Julien}}\ and\ \bibinfo {author} {\bibfnamefont {K.}~\bibnamefont {Alim}},\ }\bibfield  {title} {\bibinfo {title} {Oscillatory fluid flow drives scaling of contraction wave with system size},\ }\href@noop {} {\bibfield  {journal} {\bibinfo  {journal} {Proceedings of the National Academy of Sciences}\ }\textbf {\bibinfo {volume} {115}},\ \bibinfo {pages} {10612} (\bibinfo {year} {2018})}\BibitemShut {NoStop}%
\bibitem [{\citenamefont {Staddon}\ \emph {et~al.}(2022)\citenamefont {Staddon}, \citenamefont {Munro},\ and\ \citenamefont {Banerjee}}]{staddon2022pulsatile}%
  \BibitemOpen
  \bibfield  {author} {\bibinfo {author} {\bibfnamefont {M.~F.}\ \bibnamefont {Staddon}}, \bibinfo {author} {\bibfnamefont {E.~M.}\ \bibnamefont {Munro}},\ and\ \bibinfo {author} {\bibfnamefont {S.}~\bibnamefont {Banerjee}},\ }\bibfield  {title} {\bibinfo {title} {Pulsatile contractions and pattern formation in excitable actomyosin cortex},\ }\href@noop {} {\bibfield  {journal} {\bibinfo  {journal} {PLOS Computational Biology}\ }\textbf {\bibinfo {volume} {18}},\ \bibinfo {pages} {e1009981} (\bibinfo {year} {2022})}\BibitemShut {NoStop}%
\bibitem [{\citenamefont {Werner}\ \emph {et~al.}(2024)\citenamefont {Werner}, \citenamefont {Ray}, \citenamefont {Breen}, \citenamefont {Staddon}, \citenamefont {Jug}, \citenamefont {Banerjee},\ and\ \citenamefont {Maddox}}]{werner2024mechanical}%
  \BibitemOpen
  \bibfield  {author} {\bibinfo {author} {\bibfnamefont {M.~E.}\ \bibnamefont {Werner}}, \bibinfo {author} {\bibfnamefont {D.~D.}\ \bibnamefont {Ray}}, \bibinfo {author} {\bibfnamefont {C.}~\bibnamefont {Breen}}, \bibinfo {author} {\bibfnamefont {M.~F.}\ \bibnamefont {Staddon}}, \bibinfo {author} {\bibfnamefont {F.}~\bibnamefont {Jug}}, \bibinfo {author} {\bibfnamefont {S.}~\bibnamefont {Banerjee}},\ and\ \bibinfo {author} {\bibfnamefont {A.~S.}\ \bibnamefont {Maddox}},\ }\bibfield  {title} {\bibinfo {title} {Mechanical and biochemical feedback combine to generate complex contractile oscillations in cytokinesis},\ }\href@noop {} {\bibfield  {journal} {\bibinfo  {journal} {Current Biology}\ }\textbf {\bibinfo {volume} {34}},\ \bibinfo {pages} {3201} (\bibinfo {year} {2024})}\BibitemShut {NoStop}%
\bibitem [{\citenamefont {Picardo}\ \emph {et~al.}(2025)\citenamefont {Picardo}, \citenamefont {Jemseena},\ and\ \citenamefont {Kumar}}]{picardo2025active}%
  \BibitemOpen
  \bibfield  {author} {\bibinfo {author} {\bibfnamefont {J.~R.}\ \bibnamefont {Picardo}}, \bibinfo {author} {\bibfnamefont {V.}~\bibnamefont {Jemseena}},\ and\ \bibinfo {author} {\bibfnamefont {K.~V.}\ \bibnamefont {Kumar}},\ }\bibfield  {title} {\bibinfo {title} {Active waves from nonreciprocity and cytoplasmic exchange},\ }\href@noop {} {\bibfield  {journal} {\bibinfo  {journal} {Physical Review E}\ }\textbf {\bibinfo {volume} {112}},\ \bibinfo {pages} {L022401} (\bibinfo {year} {2025})}\BibitemShut {NoStop}%
\bibitem [{\citenamefont {Umedachi}\ \emph {et~al.}(2017)\citenamefont {Umedachi}, \citenamefont {Ito}, \citenamefont {Kobayashi}, \citenamefont {Ishiguro},\ and\ \citenamefont {Nakagaki}}]{umedachi2017response}%
  \BibitemOpen
  \bibfield  {author} {\bibinfo {author} {\bibfnamefont {T.}~\bibnamefont {Umedachi}}, \bibinfo {author} {\bibfnamefont {K.}~\bibnamefont {Ito}}, \bibinfo {author} {\bibfnamefont {R.}~\bibnamefont {Kobayashi}}, \bibinfo {author} {\bibfnamefont {A.}~\bibnamefont {Ishiguro}},\ and\ \bibinfo {author} {\bibfnamefont {T.}~\bibnamefont {Nakagaki}},\ }\bibfield  {title} {\bibinfo {title} {Response to various periods of mechanical stimuli in physarum plasmodium},\ }\href@noop {} {\bibfield  {journal} {\bibinfo  {journal} {Journal of Physics D: Applied Physics}\ }\textbf {\bibinfo {volume} {50}},\ \bibinfo {pages} {254002} (\bibinfo {year} {2017})}\BibitemShut {NoStop}%
\bibitem [{\citenamefont {{\'E}tienne}\ \emph {et~al.}(2015)\citenamefont {{\'E}tienne}, \citenamefont {Fouchard}, \citenamefont {Mitrossilis}, \citenamefont {Bufi}, \citenamefont {Durand-Smet},\ and\ \citenamefont {Asnacios}}]{etienne2015cells}%
  \BibitemOpen
  \bibfield  {author} {\bibinfo {author} {\bibfnamefont {J.}~\bibnamefont {{\'E}tienne}}, \bibinfo {author} {\bibfnamefont {J.}~\bibnamefont {Fouchard}}, \bibinfo {author} {\bibfnamefont {D.}~\bibnamefont {Mitrossilis}}, \bibinfo {author} {\bibfnamefont {N.}~\bibnamefont {Bufi}}, \bibinfo {author} {\bibfnamefont {P.}~\bibnamefont {Durand-Smet}},\ and\ \bibinfo {author} {\bibfnamefont {A.}~\bibnamefont {Asnacios}},\ }\bibfield  {title} {\bibinfo {title} {Cells as liquid motors: Mechanosensitivity emerges from collective dynamics of actomyosin cortex},\ }\href@noop {} {\bibfield  {journal} {\bibinfo  {journal} {Proceedings of the National Academy of Sciences}\ }\textbf {\bibinfo {volume} {112}},\ \bibinfo {pages} {2740} (\bibinfo {year} {2015})}\BibitemShut {NoStop}%
\bibitem [{\citenamefont {B{\"a}uerle}\ \emph {et~al.}(2017)\citenamefont {B{\"a}uerle}, \citenamefont {Kramar},\ and\ \citenamefont {Alim}}]{bauerle2017spatial}%
  \BibitemOpen
  \bibfield  {author} {\bibinfo {author} {\bibfnamefont {F.~K.}\ \bibnamefont {B{\"a}uerle}}, \bibinfo {author} {\bibfnamefont {M.}~\bibnamefont {Kramar}},\ and\ \bibinfo {author} {\bibfnamefont {K.}~\bibnamefont {Alim}},\ }\bibfield  {title} {\bibinfo {title} {Spatial mapping reveals multi-step pattern of wound healing in physarum polycephalum},\ }\href@noop {} {\bibfield  {journal} {\bibinfo  {journal} {Journal of physics D: applied physics}\ }\textbf {\bibinfo {volume} {50}},\ \bibinfo {pages} {434005} (\bibinfo {year} {2017})}\BibitemShut {NoStop}%
\bibitem [{\citenamefont {Kramar}\ and\ \citenamefont {Alim}(2021)}]{kramar2021encoding}%
  \BibitemOpen
  \bibfield  {author} {\bibinfo {author} {\bibfnamefont {M.}~\bibnamefont {Kramar}}\ and\ \bibinfo {author} {\bibfnamefont {K.}~\bibnamefont {Alim}},\ }\bibfield  {title} {\bibinfo {title} {Encoding memory in tube diameter hierarchy of living flow network},\ }\href@noop {} {\bibfield  {journal} {\bibinfo  {journal} {Proceedings of the National Academy of Sciences}\ }\textbf {\bibinfo {volume} {118}} (\bibinfo {year} {2021})}\BibitemShut {NoStop}%
\bibitem [{\citenamefont {Busson}\ \emph {et~al.}(2022)\citenamefont {Busson}, \citenamefont {Saiseau},\ and\ \citenamefont {Durand}}]{busson2022emergence}%
  \BibitemOpen
  \bibfield  {author} {\bibinfo {author} {\bibfnamefont {V.}~\bibnamefont {Busson}}, \bibinfo {author} {\bibfnamefont {R.}~\bibnamefont {Saiseau}},\ and\ \bibinfo {author} {\bibfnamefont {M.}~\bibnamefont {Durand}},\ }\bibfield  {title} {\bibinfo {title} {Emergence of dynamic contractile patterns in slime mold confined in a ring geometry},\ }\href@noop {} {\bibfield  {journal} {\bibinfo  {journal} {Journal of Physics D: Applied Physics}\ }\textbf {\bibinfo {volume} {55}},\ \bibinfo {pages} {415401} (\bibinfo {year} {2022})}\BibitemShut {NoStop}%
\bibitem [{\citenamefont {Alim}\ \emph {et~al.}(2013)\citenamefont {Alim}, \citenamefont {Amselem}, \citenamefont {Peaudecerf}, \citenamefont {Brenner},\ and\ \citenamefont {Pringle}}]{alim2013random}%
  \BibitemOpen
  \bibfield  {author} {\bibinfo {author} {\bibfnamefont {K.}~\bibnamefont {Alim}}, \bibinfo {author} {\bibfnamefont {G.}~\bibnamefont {Amselem}}, \bibinfo {author} {\bibfnamefont {F.}~\bibnamefont {Peaudecerf}}, \bibinfo {author} {\bibfnamefont {M.~P.}\ \bibnamefont {Brenner}},\ and\ \bibinfo {author} {\bibfnamefont {A.}~\bibnamefont {Pringle}},\ }\bibfield  {title} {\bibinfo {title} {Random network peristalsis in physarum polycephalum organizes fluid flows across an individual},\ }\href@noop {} {\bibfield  {journal} {\bibinfo  {journal} {Proceedings of the National Academy of Sciences}\ }\textbf {\bibinfo {volume} {110}},\ \bibinfo {pages} {13306} (\bibinfo {year} {2013})}\BibitemShut {NoStop}%
\bibitem [{\citenamefont {Fleig}\ \emph {et~al.}(2022)\citenamefont {Fleig}, \citenamefont {Kramar}, \citenamefont {Wilczek},\ and\ \citenamefont {Alim}}]{fleig2022emergence}%
  \BibitemOpen
  \bibfield  {author} {\bibinfo {author} {\bibfnamefont {P.}~\bibnamefont {Fleig}}, \bibinfo {author} {\bibfnamefont {M.}~\bibnamefont {Kramar}}, \bibinfo {author} {\bibfnamefont {M.}~\bibnamefont {Wilczek}},\ and\ \bibinfo {author} {\bibfnamefont {K.}~\bibnamefont {Alim}},\ }\bibfield  {title} {\bibinfo {title} {Emergence of behaviour in a self-organized living matter network},\ }\href@noop {} {\bibfield  {journal} {\bibinfo  {journal} {Elife}\ }\textbf {\bibinfo {volume} {11}},\ \bibinfo {pages} {e62863} (\bibinfo {year} {2022})}\BibitemShut {NoStop}%
\bibitem [{\citenamefont {Kuroda}\ \emph {et~al.}(2015)\citenamefont {Kuroda}, \citenamefont {Takagi}, \citenamefont {Nakagaki},\ and\ \citenamefont {Ueda}}]{kuroda2015allometry}%
  \BibitemOpen
  \bibfield  {author} {\bibinfo {author} {\bibfnamefont {S.}~\bibnamefont {Kuroda}}, \bibinfo {author} {\bibfnamefont {S.}~\bibnamefont {Takagi}}, \bibinfo {author} {\bibfnamefont {T.}~\bibnamefont {Nakagaki}},\ and\ \bibinfo {author} {\bibfnamefont {T.}~\bibnamefont {Ueda}},\ }\bibfield  {title} {\bibinfo {title} {Allometry in physarum plasmodium during free locomotion: size versus shape, speed and rhythm},\ }\href@noop {} {\bibfield  {journal} {\bibinfo  {journal} {Journal of experimental biology}\ }\textbf {\bibinfo {volume} {218}},\ \bibinfo {pages} {3729} (\bibinfo {year} {2015})}\BibitemShut {NoStop}%
\bibitem [{\citenamefont {Oettmeier}\ and\ \citenamefont {D{\"o}bereiner}(2019)}]{oettmeier2019lumped}%
  \BibitemOpen
  \bibfield  {author} {\bibinfo {author} {\bibfnamefont {C.}~\bibnamefont {Oettmeier}}\ and\ \bibinfo {author} {\bibfnamefont {H.-G.}\ \bibnamefont {D{\"o}bereiner}},\ }\bibfield  {title} {\bibinfo {title} {A lumped parameter model of endoplasm flow in physarum polycephalum explains migration and polarization-induced asymmetry during the onset of locomotion},\ }\href@noop {} {\bibfield  {journal} {\bibinfo  {journal} {PloS one}\ }\textbf {\bibinfo {volume} {14}},\ \bibinfo {pages} {e0215622} (\bibinfo {year} {2019})}\BibitemShut {NoStop}%
\bibitem [{\citenamefont {Alim}\ \emph {et~al.}(2017)\citenamefont {Alim}, \citenamefont {Andrew}, \citenamefont {Pringle},\ and\ \citenamefont {Brenner}}]{alim2017mechanism}%
  \BibitemOpen
  \bibfield  {author} {\bibinfo {author} {\bibfnamefont {K.}~\bibnamefont {Alim}}, \bibinfo {author} {\bibfnamefont {N.}~\bibnamefont {Andrew}}, \bibinfo {author} {\bibfnamefont {A.}~\bibnamefont {Pringle}},\ and\ \bibinfo {author} {\bibfnamefont {M.~P.}\ \bibnamefont {Brenner}},\ }\bibfield  {title} {\bibinfo {title} {Mechanism of signal propagation in physarum polycephalum},\ }\href@noop {} {\bibfield  {journal} {\bibinfo  {journal} {Proceedings of the National Academy of Sciences}\ }\textbf {\bibinfo {volume} {114}},\ \bibinfo {pages} {5136} (\bibinfo {year} {2017})}\BibitemShut {NoStop}%
\bibitem [{\citenamefont {Mayne}\ \emph {et~al.}(2017)\citenamefont {Mayne}, \citenamefont {Jones}, \citenamefont {Gale},\ and\ \citenamefont {Adamatzky}}]{mayne2017coupled}%
  \BibitemOpen
  \bibfield  {author} {\bibinfo {author} {\bibfnamefont {R.}~\bibnamefont {Mayne}}, \bibinfo {author} {\bibfnamefont {J.}~\bibnamefont {Jones}}, \bibinfo {author} {\bibfnamefont {E.}~\bibnamefont {Gale}},\ and\ \bibinfo {author} {\bibfnamefont {A.}~\bibnamefont {Adamatzky}},\ }\bibfield  {title} {\bibinfo {title} {On coupled oscillator dynamics and incident behaviour patterns in slime mould physarum polycephalum: emergence of wave packets, global streaming clock frequencies and anticipation of periodic stimuli},\ }\href@noop {} {\bibfield  {journal} {\bibinfo  {journal} {International Journal of Parallel, Emergent and Distributed Systems}\ }\textbf {\bibinfo {volume} {32}},\ \bibinfo {pages} {95} (\bibinfo {year} {2017})}\BibitemShut {NoStop}%
\bibitem [{\citenamefont {Marbach}\ \emph {et~al.}(2023)\citenamefont {Marbach}, \citenamefont {Ziethen}, \citenamefont {Bastin}, \citenamefont {B{\"a}uerle},\ and\ \citenamefont {Alim}}]{marbach2023vein}%
  \BibitemOpen
  \bibfield  {author} {\bibinfo {author} {\bibfnamefont {S.}~\bibnamefont {Marbach}}, \bibinfo {author} {\bibfnamefont {N.}~\bibnamefont {Ziethen}}, \bibinfo {author} {\bibfnamefont {L.}~\bibnamefont {Bastin}}, \bibinfo {author} {\bibfnamefont {F.~K.}\ \bibnamefont {B{\"a}uerle}},\ and\ \bibinfo {author} {\bibfnamefont {K.}~\bibnamefont {Alim}},\ }\bibfield  {title} {\bibinfo {title} {Vein fate determined by flow-based but time-delayed integration of network architecture},\ }\href@noop {} {\bibfield  {journal} {\bibinfo  {journal} {Elife}\ }\textbf {\bibinfo {volume} {12}},\ \bibinfo {pages} {e78100} (\bibinfo {year} {2023})}\BibitemShut {NoStop}%
\bibitem [{\citenamefont {Nakagaki}\ \emph {et~al.}(2000)\citenamefont {Nakagaki}, \citenamefont {Yamada},\ and\ \citenamefont {Ueda}}]{nakagaki2000interaction}%
  \BibitemOpen
  \bibfield  {author} {\bibinfo {author} {\bibfnamefont {T.}~\bibnamefont {Nakagaki}}, \bibinfo {author} {\bibfnamefont {H.}~\bibnamefont {Yamada}},\ and\ \bibinfo {author} {\bibfnamefont {T.}~\bibnamefont {Ueda}},\ }\bibfield  {title} {\bibinfo {title} {Interaction between cell shape and contraction pattern in the physarum plasmodium},\ }\href@noop {} {\bibfield  {journal} {\bibinfo  {journal} {Biophysical chemistry}\ }\textbf {\bibinfo {volume} {84}},\ \bibinfo {pages} {195} (\bibinfo {year} {2000})}\BibitemShut {NoStop}%
\bibitem [{\citenamefont {Yoshimoto}\ \emph {et~al.}(1981)\citenamefont {Yoshimoto}, \citenamefont {Matsumura},\ and\ \citenamefont {Kamiya}}]{yoshimoto1981simultaneous}%
  \BibitemOpen
  \bibfield  {author} {\bibinfo {author} {\bibfnamefont {Y.}~\bibnamefont {Yoshimoto}}, \bibinfo {author} {\bibfnamefont {F.}~\bibnamefont {Matsumura}},\ and\ \bibinfo {author} {\bibfnamefont {N.}~\bibnamefont {Kamiya}},\ }\bibfield  {title} {\bibinfo {title} {Simultaneous oscillations of ca2+ efflux and tension generation in the permealized plasmodial strand of physarum},\ }\href@noop {} {\bibfield  {journal} {\bibinfo  {journal} {Cell motility}\ }\textbf {\bibinfo {volume} {1}},\ \bibinfo {pages} {433} (\bibinfo {year} {1981})}\BibitemShut {NoStop}%
\bibitem [{\citenamefont {Saiseau}\ \emph {et~al.}(2024)\citenamefont {Saiseau}, \citenamefont {Busson}, \citenamefont {X{\'e}nard},\ and\ \citenamefont {Durand}}]{saiseau2024network}%
  \BibitemOpen
  \bibfield  {author} {\bibinfo {author} {\bibfnamefont {R.}~\bibnamefont {Saiseau}}, \bibinfo {author} {\bibfnamefont {V.}~\bibnamefont {Busson}}, \bibinfo {author} {\bibfnamefont {L.}~\bibnamefont {X{\'e}nard}},\ and\ \bibinfo {author} {\bibfnamefont {M.}~\bibnamefont {Durand}},\ }\bibfield  {title} {\bibinfo {title} {Network emergence and reorganization in confined slime moulds},\ }\href@noop {} {\bibfield  {journal} {\bibinfo  {journal} {Journal of Physics D: Applied Physics}\ }\textbf {\bibinfo {volume} {57}},\ \bibinfo {pages} {145401} (\bibinfo {year} {2024})}\BibitemShut {NoStop}%
\bibitem [{\citenamefont {Zhang}\ \emph {et~al.}(2017)\citenamefont {Zhang}, \citenamefont {Guy}, \citenamefont {Lasheras},\ and\ \citenamefont {Del~{\'A}lamo}}]{zhang2017self}%
  \BibitemOpen
  \bibfield  {author} {\bibinfo {author} {\bibfnamefont {S.}~\bibnamefont {Zhang}}, \bibinfo {author} {\bibfnamefont {R.~D.}\ \bibnamefont {Guy}}, \bibinfo {author} {\bibfnamefont {J.~C.}\ \bibnamefont {Lasheras}},\ and\ \bibinfo {author} {\bibfnamefont {J.~C.}\ \bibnamefont {Del~{\'A}lamo}},\ }\bibfield  {title} {\bibinfo {title} {Self-organized mechano-chemical dynamics in amoeboid locomotion of physarum fragments},\ }\href@noop {} {\bibfield  {journal} {\bibinfo  {journal} {Journal of physics D: Applied physics}\ }\textbf {\bibinfo {volume} {50}},\ \bibinfo {pages} {204004} (\bibinfo {year} {2017})}\BibitemShut {NoStop}%
\bibitem [{\citenamefont {Pikovsky}\ \emph {et~al.}(2001)\citenamefont {Pikovsky}, \citenamefont {Rosenblum},\ and\ \citenamefont {Kurths}}]{Pikovsky_Rosenblum_Kurths_2001}%
  \BibitemOpen
  \bibfield  {author} {\bibinfo {author} {\bibfnamefont {A.}~\bibnamefont {Pikovsky}}, \bibinfo {author} {\bibfnamefont {M.}~\bibnamefont {Rosenblum}},\ and\ \bibinfo {author} {\bibfnamefont {J.}~\bibnamefont {Kurths}},\ }\href@noop {} {\emph {\bibinfo {title} {Synchronization: A Universal Concept in Nonlinear Sciences}}},\ Cambridge Nonlinear Science Series\ (\bibinfo  {publisher} {Cambridge University Press},\ \bibinfo {year} {2001})\BibitemShut {NoStop}%
\bibitem [{\citenamefont {Wilkinson}\ \emph {et~al.}(2023)\citenamefont {Wilkinson}, \citenamefont {Koziol}, \citenamefont {Alim},\ and\ \citenamefont {Roper}}]{wilkinson2023flow}%
  \BibitemOpen
  \bibfield  {author} {\bibinfo {author} {\bibfnamefont {R.}~\bibnamefont {Wilkinson}}, \bibinfo {author} {\bibfnamefont {M.}~\bibnamefont {Koziol}}, \bibinfo {author} {\bibfnamefont {K.}~\bibnamefont {Alim}},\ and\ \bibinfo {author} {\bibfnamefont {M.}~\bibnamefont {Roper}},\ }\bibfield  {title} {\bibinfo {title} {Flow modes provide a quantification of physarum network peristalsis},\ }\href@noop {} {\bibfield  {journal} {\bibinfo  {journal} {Fungal Ecology}\ }\textbf {\bibinfo {volume} {65}},\ \bibinfo {pages} {101283} (\bibinfo {year} {2023})}\BibitemShut {NoStop}%
\bibitem [{\citenamefont {Nishikawa}\ \emph {et~al.}(2017)\citenamefont {Nishikawa}, \citenamefont {Naganathan}, \citenamefont {J{\"u}licher},\ and\ \citenamefont {Grill}}]{nishikawa2017controlling}%
  \BibitemOpen
  \bibfield  {author} {\bibinfo {author} {\bibfnamefont {M.}~\bibnamefont {Nishikawa}}, \bibinfo {author} {\bibfnamefont {S.~R.}\ \bibnamefont {Naganathan}}, \bibinfo {author} {\bibfnamefont {F.}~\bibnamefont {J{\"u}licher}},\ and\ \bibinfo {author} {\bibfnamefont {S.~W.}\ \bibnamefont {Grill}},\ }\bibfield  {title} {\bibinfo {title} {Controlling contractile instabilities in the actomyosin cortex},\ }\href@noop {} {\bibfield  {journal} {\bibinfo  {journal} {Elife}\ }\textbf {\bibinfo {volume} {6}},\ \bibinfo {pages} {e19595} (\bibinfo {year} {2017})}\BibitemShut {NoStop}%
\end{thebibliography}%

\end{document}